%
\let\useblackboard=\iftrue
%
%
\newfam\black

\input harvmac
\input epsf.tex

\useblackboard
\message{If you do not have msbm (blackboard bold) fonts,}
\message{change the option at the top of the tex file.}
\font\blackboard=msbm10
\font\blackboards=msbm7
\font\blackboardss=msbm5

\font\hesmallrm = cmr5 at 3truept
\font\hsmallrm=cmr7
\textfont\black=\blackboard
\scriptfont\black=\blackboards
\scriptscriptfont\black=\blackboardss

\else

\fi
\def\boxit#1{\vbox{\hrule\hbox{\vrule\kern8pt
\vbox{\hbox{\kern8pt}\hbox{\vbox{#1}}\hbox{\kern8pt}}
\kern8pt\vrule}\hrule}}
\def\mathboxit#1{\vbox{\hrule\hbox{\vrule\kern8pt\vbox{\kern8pt
\hbox{$\displaystyle #1$}\kern8pt}\kern8pt\vrule}\hrule}}

\def\subsubsec#1{\ifnum\lastpenalty>9000\else\bigbreak\fi
\noindent{\it{#1}}\par\nobreak\medskip\nobreak}
\def\yboxit#1#2{\vbox{\hrule height #1 \hbox{\vrule width #1
\vbox{#2}\vrule width #1 }\hrule height #1 }}
\def\fillbox#1{\hbox to #1{\vbox to #1{\vfil}\hfil}}
\def\ybox{{\lower 1.3pt \yboxit{0.4pt}{\fillbox{8pt}}\hskip-0.2pt}}

\

\def\drawbox#1#2{\hrule height#2pt 
        \hbox{\vrule width#2pt height#1pt \kern#1pt \vrule width#2pt}
              \hrule height#2pt}

\def\Fund#1#2{\vcenter{\vbox{\drawbox{#1}{#2}}}}
\def\Asym#1#2{\vcenter{\vbox{\drawbox{#1}{#2}
              \kern-#2pt       
              \drawbox{#1}{#2}}}}

\def\sfund{\Fund{3.25}{0.2}}
\def\sasym{\Asym{3.25}{0.2}}


\def\frac#1#2{{#1 \over #2}}




\lref\AffleckXZ{
  I.~Affleck, M.~Dine and N.~Seiberg,
  ``Dynamical Supersymmetry Breaking In Four-Dimensions And Its
  Phenomenological Implications,''
  Nucl.\ Phys.\  B {\bf 256}, 557 (1985).
}

\lref\DineAG{
  M.~Dine, A.~E.~Nelson, Y.~Nir and Y.~Shirman,
  ``New tools for low-energy dynamical supersymmetry breaking,''
  Phys.\ Rev.\  D {\bf 53}, 2658 (1996)
  [arXiv:hep-ph/9507378].
}

\lref\WittenFP{
  E.~Witten,
  ``An SU(2) anomaly,''
  Phys.\ Lett.\  B {\bf 117}, 324 (1982).
}

\lref\SeibergQJ{
  N.~Seiberg, T.~Volansky and B.~Wecht,
  ``Semi-direct Gauge Mediation,''
  JHEP {\bf 0811}, 004 (2008)
  [arXiv:0809.4437 [hep-ph]].
}

\lref\GrisaruVE{
  M.~T.~Grisaru, M.~Rocek and R.~von Unge,
  ``Effective K\"ahler Potentials,''
  Phys.\ Lett.\  B {\bf 383}, 415 (1996)
  [arXiv:hep-th/9605149].
}

\lref\IntriligatorDD{
  K.~A.~Intriligator, N.~Seiberg and D.~Shih,
  ``Dynamical SUSY breaking in meta-stable vacua,''
  JHEP {\bf 0604}, 021 (2006)
  [arXiv:hep-th/0602239].
}

\lref\PoppitzVD{
  E.~Poppitz and S.~P.~Trivedi,
  ``Dynamical supersymmetry breaking,''
  Ann.\ Rev.\ Nucl.\ Part.\ Sci.\  {\bf 48}, 307 (1998)
  [arXiv:hep-th/9803107].
}

\lref\DineZA{
  M.~Dine, W.~Fischler and M.~Srednicki,
  ``Supersymmetric Technicolor,''
  Nucl.\ Phys.\  B {\bf 189}, 575 (1981).
}

\lref\DimopoulosAU{
  S.~Dimopoulos and S.~Raby,
  ``Supercolor,''
  Nucl.\ Phys.\  B {\bf 192}, 353 (1981).
}

\lref\DineGU{
  M.~Dine and W.~Fischler,
  ``A Phenomenological Model Of Particle Physics Based On Supersymmetry,''
  Phys.\ Lett.\  B {\bf 110}, 227 (1982).
}

\lref\NappiHM{
  C.~R.~Nappi and B.~A.~Ovrut,
  ``Supersymmetric Extension Of The SU(3) X SU(2) X U(1) Model,''
  Phys.\ Lett.\  B {\bf 113}, 175 (1982).
}

\lref\DineZB{
  M.~Dine and W.~Fischler,
  ``A Supersymmetric Gut,''
  Nucl.\ Phys.\  B {\bf 204}, 346 (1982).
}

\lref\AlvarezGaumeWY{
  L.~Alvarez-Gaume, M.~Claudson and M.~B.~Wise,
  ``Low-Energy Supersymmetry,''
  Nucl.\ Phys.\  B {\bf 207}, 96 (1982).
}

\lref\DimopoulosGM{
  S.~Dimopoulos and S.~Raby,
  ``Geometric Hierarchy,''
  Nucl.\ Phys.\  B {\bf 219}, 479 (1983).
}

\lref\ArkaniHamedJV{
  N.~Arkani-Hamed, J.~March-Russell and H.~Murayama,
  ``Building models of gauge-mediated supersymmetry breaking without a
  messenger sector,''
  Nucl.\ Phys.\  B {\bf 509}, 3 (1998)
  [arXiv:hep-ph/9701286].
}

\lref\PoppitzFW{
  E.~Poppitz and S.~P.~Trivedi,
  ``New models of gauge and gravity mediated supersymmetry breaking,''
  Phys.\ Rev.\  D {\bf 55}, 5508 (1997)
  [arXiv:hep-ph/9609529].
}

\lref\MurayamaPB{
  H.~Murayama,
  ``A model of direct gauge mediation,''
  Phys.\ Rev.\ Lett.\  {\bf 79}, 18 (1997)
  [arXiv:hep-ph/9705271].
}

\lref\DineI{
  M.~Dine and A.~E.~Nelson,
  ``Dynamical supersymmetry breaking at low-energies,''
  Phys.\ Rev.\  D {\bf 48}, 1277 (1993)
  [arXiv:hep-ph/9303230].
}

\lref\DineII{
  M.~Dine, A.~E.~Nelson and Y.~Shirman,
  ``Low-Energy Dynamical Supersymmetry Breaking Simplified,''
  Phys.\ Rev.\  D {\bf 51}, 1362 (1995)
  [arXiv:hep-ph/9408384].
}

\lref\MeadeWD{
  P.~Meade, N.~Seiberg and D.~Shih,
  ``General Gauge Mediation,''
  arXiv:0801.3278 [hep-ph].
}

\lref\KomargodskiPC{
  Z.~Komargodski and N.~Seiberg,
  ``Comments on the Fayet-Iliopoulos Term in Field Theory and Supergravity,''
  arXiv:0904.1159 [hep-th].
}

\lref\BaggerHH{
  J.~Bagger, E.~Poppitz and L.~Randall,
  ``The R axion from dynamical supersymmetry breaking,''
  Nucl.\ Phys.\  B {\bf 426}, 3 (1994)
  [arXiv:hep-ph/9405345].
}

\lref\CarpenterRJ{
  L.~M.~Carpenter,
  ``Gauge Mediation with D-terms,''
  arXiv:0809.0026 [hep-ph].
}

\lref\PoppitzFH{
  E.~Poppitz and S.~P.~Trivedi,
  ``Some examples of chiral moduli spaces and dynamical supersymmetry
  breaking,''
  Phys.\ Lett.\  B {\bf 365}, 125 (1996)
  [arXiv:hep-th/9507169].
}

\lref\ShadmiJY{
  Y.~Shadmi and Y.~Shirman,
  ``Dynamical supersymmetry breaking,''
  Rev.\ Mod.\ Phys.\  {\bf 72}, 25 (2000)
  [arXiv:hep-th/9907225].
}

\lref\PoppitzXW{
  E.~Poppitz and S.~P.~Trivedi,
  ``Some remarks on gauge-mediated supersymmetry breaking,''
  Phys.\ Lett.\  B {\bf 401}, 38 (1997)
  [arXiv:hep-ph/9703246].
}

\lref\KomargodskiJF{
  Z.~Komargodski and D.~Shih,
  ``Notes on SUSY and R-Symmetry Breaking in Wess-Zumino Models,''
  arXiv:0902.0030 [hep-th].
}

\lref\IbeWP{
  M.~Ibe, Y.~Nakayama and T.~T.~Yanagida,
  ``Conformal gauge mediation,''
  Phys.\ Lett.\  B {\bf 649}, 292 (2007)
  [arXiv:hep-ph/0703110].
}

\lref\PoppitzTX{
  E.~Poppitz and L.~Randall,
  ``Low-energy Kahler potentials in supersymmetric gauge theories with (almost)
  flat directions,''
  Phys.\ Lett.\  B {\bf 336}, 402 (1994)
  [arXiv:hep-th/9407185].
}

\Title{\vbox{\baselineskip12pt \hbox{UUITP-12/09}}}
{\vbox{\centerline{Semi-Direct Gauge Mediation with the 4-1 Model}}}
\centerline{Henriette Elvang\foot{{\hsmallrm On leave of absence from Uppsala University.}} and Brian Wecht}
\bigskip
\centerline{{\it School of Natural Sciences, Institute for Advanced Study, Princeton, NJ 08540, USA}} 
\bigskip\bigskip

\noindent
We analyze a model of Semi-Direct Gauge Mediation in which the hidden sector is the 4-1 model and the messenger fields are charged under the $U(1)$ gauge group. 
At tree level, the SUSY-breaking F-terms induce D-terms from which SUSY-split messenger masses arise. We calculate these masses by three complementary methods. Additionally, we compute the one-loop corrections to the masses. We consider this model both with and without a Fayet-Iliopoulos term for the hidden sector $U(1)$. Finally, we write down a simple model of Minimal Gauge Mediation in which the only scale is dynamically generated.

\smallskip
\Date{April 2009}

\newsec{Introduction and summary}
\seclab\intro
 
Supersymmetry (SUSY) is one of the most exciting candidates for physics beyond the Standard Model. However, finding realistic models of SUSY breaking turns out to be an exceptionally difficult problem. In particular, 
SUSY must be broken in a hidden sector, so a crucial ingredient is how SUSY breaking is mediated to the supersymmetric Standard Model (SSM). The two most studied frameworks are gravity mediation and gauge mediation. In either scenario, the models tend to have problems which present difficulties for phenomenology. This motivates the study of new models and frameworks for mediating SUSY breaking. 

Models in which SUSY breaking is communicated to the SSM via gauge interactions have the advantage of avoiding large flavor-violating effects, which are difficult to suppress in other models. The first such gauge-mediated models were presented in \refs{\DineZA\DimopoulosAU\DineGU
\NappiHM\DineZB\AlvarezGaumeWY-\DimopoulosGM}, and recently a very general description of the framework of gauge mediation appeared in \MeadeWD. The two types of gauge mediation which have seen the most attention are {\sl Direct Mediation} and {\sl Minimal Gauge Mediation}. In Direct Mediation models \refs{\PoppitzFW\ArkaniHamedJV-\MurayamaPB}, the SSM gauge group is embedded in a weakly gauged flavor symmetry of the hidden sector. In Minimal Gauge Mediation models \refs{\DineI\DineII-\DineAG}, there is a messenger sector which communicates with the SSM through gauge interactions, and couples to the hidden sector via a tree-level superpotential.  

{\sl Semi-Direct Gauge Mediation} is a synthesis of these two frameworks. In this scenario, the Standard Model  gauge group is embedded in a (weakly gauged) flavor symmetry of the messenger sector, but the messengers only communicate with the hidden sector via gauge interactions. 
This setup was proposed in the recent work \SeibergQJ. A similar idea was studied in \IbeWP. 
In  \SeibergQJ, the primary features of Semi-Direct Gauge Mediation were illustrated using as the hidden sector 
the ``3-2 model'' of dynamical SUSY breaking \AffleckXZ. This model has a $SU(3)\times SU(2)$ gauge group, and \SeibergQJ\ added $SU(2)$ doublet messenger fields.  
It was found that SUSY breaking in the hidden sector induced SUSY-split messenger masses, which could be calculated using three complementary methods.

In this work, we extend the program of Semi-Direct Gauge Mediation by taking the hidden sector to be another calculable model of dynamical SUSY breaking, namely the ``4-1 model" first presented in \refs{\DineAG,\PoppitzFH}. The model has a $SU(4)\times U(1)$ gauge group and four matter fields.
The matter fields are all charged under the $U(1)$, and  under the $SU(4)$ they are a singlet, fundamental, anti-fundamental and antisymmetric two-index tensor. 
The superpotential includes a tree level term as well as a dynamically generated contribution. The dynamics and supersymmetry breaking in this model have
been studied by several authors,\foot{See \refs{\DineAG,\PoppitzFH\PoppitzVD-\ShadmiJY} and references therein. Also, the 4-1 model was recently discussed in the context of General Gauge Mediation in \CarpenterRJ.} but for completeness we rederive in this paper many of the relevant properties. See in particular Section 2.

In Section 3 we couple pairs of $U(1)$-charged messenger fields $L$ and $\bar L$ to the 4-1 model. There are no superpotential couplings between the messengers and the hidden sector; they interact only through the gauge interactions. At tree level, the messengers feel SUSY breaking via  ``diagonal'' mass terms $m_d^2 L L^\dagger$. We calculate $m_d^2$ explicitly, using three complimentary methods: 
\smallskip
\noindent ~$\bullet$ microscopic analysis using the fundamental fields and Wess-Zumino gauge (Section 3.1).

\noindent ~$\bullet$ macroscopic analysis in terms of composite gauge invariant operators (Section 3.2). 

\noindent ~$\bullet$ low-energy effective theory in unitary gauge (Section 3.3). 
\smallskip

\noindent The first method is the most straightforward calculation. The two other methods serve not only as checks but also illuminate interesting aspects of the result. 

One example of this is that the mass $m_d^2$ turns out to be independent of the $U(1)$ gauge coupling $g_1$. This is a surprising result from the point of view of the microscopic calculation. After all, the messengers interact with the hidden sector only through gauge interactions! However, in the gauge-invariant description, the Higgsed gauge fields have been integrated out and the result is an effective non-linear sigma model which 
is independent of $g_1$. In this picture, the mass splittings $m_d^2$ arise from the curvature on moduli space and therefore cannot depend on $g_1$.

The complimentary approaches also clarify the relationship between F-terms and D-terms.
One of the most surprising effects is that the SUSY-breaking F-terms induce tree-level D-terms, from which the diagonal masses $m_d^2$ arise. 
The precise connection between the D-terms and the F-terms follows from the calculation in unitary gauge. In this calculation, the gauge fields are integrated out by solving their equation of motion to leading order. One finds that the superfield equation of motion implies a relationship between the D-term of the vectors and the F-terms $\cal F$ of the chiral fields, $D \sim |{\cal  F}|^2/|\phi|^2$. The SUSY-split diagonal masses $m_d^2$ are proportional to this term. This is a generic feature of Semi-Direct models, and is true even when the hidden sector gauge group has no Abelian factors. This stands in contrast to MGM models, in which such masses show up via FI terms. 

Classically, SUSY breaking only generates diagonal masses for the messengers. When the one-loop correction to the K\"ahler potential is included (Section 3.4), off-diagonal terms $m_{od}^2 L \bar L$ are generated and corrections to the diagonal masses now make the supertrace over the messenger sector non-vanishing. We show that ${\rm Str}\,m_{\rm msg}^2 < 0$.

Because the 4-1 model gauge group includes a $U(1)$ factor, we are free to add a Fayet-Iliopoulos term $\xi$. The effects of this are studied in Section 4. The off-diagonal masses turn out to be bounded as functions of $\xi$, and it may therefore be possible to 
keep the off-diagonal terms small, while making ${\rm Str}\,m_{\rm msg}^2$  large (and negative). 
This may be useful for achieving $m_{\tilde f}^2 > 0$ for SSM sfermions $\tilde f$.

In Section 5 we take a step back to study the consistency of the model. In particular we clarify the regime in parameter space in which our calculations are valid. As in the Semi-Direct 3-2 model  \SeibergQJ , we have to include an explicit superpotential  mass term for the messengers. The corresponding mass parameter $m$ must be large enough to ensure that the messenger fields do not get non-vanishing vevs, but in order to be relevant in the effective low-energy theory, $m$ should not be larger than the masses of the Higgsed vector fields. We determine the precise conditions for this in Section 5 and derive the needed constraints on the parameters, including $\xi$. 

Having an explicit mass term 
for the messengers
is somewhat undesirable. In Section 5.3 we modify the Semi-Direct model to become a simple model of Minimal Gauge Mediation with a meta-stable vacuum. In this model, $m=0$ and the only scale is the dynamically generated scale of the 4-1 model. 

We conclude the paper in Section 6 with brief preliminary comments on the phenomenology of the 4-1 Semi-Direct model. 
Semi-Direct models fit within the framework of General Gauge Mediation \MeadeWD , but specific models can have phenomenology which is not captured by the unified description. In our work, one new feature is the explicit dependence on the FI parameter.

Two appendices collect technical material: in Appendix A, we determine the D-flat directions of the 4-1 model, and in Appendix B we write down the generators of $SU(4)$ which are needed for the unitary gauge calculations.


\newsec{4-1 model}
\seclab\fourone

The 4-1 model has gauge group $SU(4) \times U(1)$ and matter content
\eqn\matter{
\matrix{ & SU(4) & U(1) & U(1)_R \cr
S & {\bf 1} & 4 &6 \cr
F_i & {\bf 4} & -3 & 0  \cr
{\bar F}^i & \bar {\bf 4} & -1 & -4 \cr
A_{ij} & {\bf 6} & 2 &0 }
}
In the last column, we have made a convenient assignment of charges for the global non-anomalous $U(1)_R$ symmetry. 

Let us first consider the D-flat moduli space.
The general solution to the $SU(4)$ D-flatness conditions is 
\eqn\aisFin{
A = \frac{a}{\sqrt{2}} \left ( \matrix{ i\,\sigma_2 & 0 \cr 0 & i \, \sigma_2} \right )\,,\quad~~
F = \bar F^T=  \left ( \matrix{ b  \cr 0 \cr 0 \cr 0} \right ) \, ,\quad ~~
 S= c\, e^{i\phi_c} \, , 
}
where $a,b,c$ are real positive numbers. The derivation of \aisFin\ is outlined in Appendix A. A further restriction is imposed by the $U(1)$ D-flatness condition,
\eqn\uonedflat{
2a^2 - 4b^2 + 4c^2 = 0.
}  

The moduli space can be parametrized by two independent gauge-invariant operators, $B \equiv S \bar F^i F_i$ and $Y \equiv {1 \over 4} \bar F^i F_i {\rm Pf} A$. (Equivalently, we could write $Y=\bar F^i F_j A_{ik} A_{lm} \epsilon^{jklm}$.)
At a generic point on moduli space, there is an unbroken $SU(2)$ gauge group.

We add to this model the tree-level superpotential
\eqn\Wtree{
  W_{\rm tree} = h\,S \bar F^i F_i = hB.
}
The F-terms force $b=0$. Hence $a=c=0$ by \uonedflat, and the superpotential thus removes both classical flat directions. 

The unbroken $SU(2)$ undergoes gaugino condensation, which generates a superpotential $\tilde{W}_{\rm dyn}  = 2 \Lambda_2^3$. In the original $SU(4)$ theory, the dynamically generated superpotential is $W_{\rm dyn}  = 2 \left ( \Lambda_4^{10} / Y\right)^{1/2}$. 
One can motivate that $\tilde{W}_{\rm dyn} \sim W_{\rm dyn}$ via symmetries, since it is the only possible consistent term. To see how $W_{\rm dyn}$ is produced from the $SU(2)$ answer (ignoring numerical factors), we can go to a point on moduli space where $b \gg a$ \DineAG. There, 
$SU(4)$ is broken to $SU(3)$ with a massless ${\bf 3} + \bar{\bf 3}$. Matching dynamical scales at the scale $b$ gives $\Lambda_4^{10} = b^2 \Lambda_3^8$. Next, at the scale $a$, 
$SU(3)$ is broken to $SU(2)$ with singlets,
so $\Lambda_3^8 = a^2 \Lambda_2^6$. Thus $\Lambda_4^{10}=a^2b^2 \Lambda_2^6 = \langle Y \rangle \Lambda_2^6$. This reproduces the dynamically generated superpotential $ \left ( \Lambda_4^{10} / Y \right )^{1/2}$ from gaugino condensation in the $SU(2)$ theory. 
Henceforth we set $\Lambda \equiv \Lambda_4$.

Including the dynamically generated term $W_{\rm dyn}$ 
the full superpotential for the $SU(4)$ theory is then
\eqn\Wdyn{
W
= h \, B + 2 \left ( {\Lambda^{10} \over Y} \right )^{1/2}.
}
As we will see in more detail below, the superpotential \Wdyn\ breaks supersymmetry.

Before minimizing the scalar potential, it is convenient to remove the dynamical scale $\Lambda$ and work in terms of dimensionless quantities. This is done by rescaling all fields $\phi \to \Lambda\, h^{-1/5} \phi$ and gives
\eqn\VtotA{
V = \left ( V_F + \frac{1}{\epsilon_1} V_{D1}
+ \frac{1}{\epsilon_4} V_{D4} \right ) \, h^{6/5}\, \Lambda^4\,,
}
with 
\eqn\VtotAb{
V_F = |\partial W |^2 \, , \quad\quad
V_{D1} = {1 \over 8} D_{U(1)}^2\, ,\quad\quad
V_{D4} = {1 \over 8} (D^a_{SU(4)})^2
\quad~~ {\rm and}\quad ~~\epsilon_{1,4} = \frac{h^2}{g_{1,4}^{2}}
.
}
Explicit expressions for the D-terms are given in Appendix A.2. 
 
We assume that $h \ll g_1 \ll g_4 \ll 1$. 
The hierarchy between the gauge couplings is automatic because the $U(1)$ is IR free.
In this limit, the vevs are large and the theory is calculable. Since $\epsilon \ll 1$ we can minimize the potential to leading order in $\epsilon$  by minimizing $V_F$ on the D-flat directions. 
Imposing \aisFin\ gives
\eqn\VF{
V_F =  h^{6/5} \Lambda^4 
\left(
2 \, \left | b c\, e^{i\phi_c} - \frac{1}{a b^2} \right |^2
 +b^4 + \frac{4}{a^4 b^2} 
\right) \,.
}
This is minimized for $\phi_c=0$.  Extremizing with respect to the remaining fields, we find that the minimum is located at
\eqn\minA{
(a,b,c) = (1.492, 1.102, 0.318)\, ,
~~~~~~~~
V_{\rm min} = 2.22 \,  h^{6/5} \Lambda^4 \,.
}

We now perturb around the minimum \minA\ to find the $O(\epsilon)$-correction. 
This correction makes the D-terms nonzero. We use $\phi^{(0)}$ to denote the solution \aisFin\ with the values \minA. Writing $\phi = \phi^{(0)}+ \epsilon\, \phi^{(1)}$, we find
\eqn\Vexp{
V(\phi) = V_F(\phi^{(0)}) 
+ \epsilon\,
 \Big( \partial_{\phi_A} V_F \Big|_{\phi=\phi^{(0)}} \, \phi_A^{(1)}
+ \frac{1}{2} \partial_{\phi_A} \partial_{\phi_B } V_D \Big|_{\phi=\phi^{(0)}}
 \, \phi^{(1)}_A\, \phi^{(1)}_B
 \Big)+ O(\epsilon^2) \, .
}
We work in terms of real variables, so $\phi$ is a 30-component vector and $A,B = 1,\dots,30$.
To determine $\phi^{(1)}$ such that the $O(\epsilon)$-term is minimized, we must solve the linear system
\eqn\tos{
  \partial_{\phi_A} V_F \Big|_{\phi=\phi^{(0)}} 
+ \partial_{\phi_A} \partial_{\phi_B } V_D \Big|_{\phi=\phi^{(0)}}
 \, \phi^{(1)}_B =0 \, .
}
The first term in this equation, $\partial V_F$, is only non-vanishing in the directions where $\phi^{(0)}$ is nonzero. The second term involves a rank 13 matrix. One can straightforwardly solve this equation for these 13 fields in terms of the remaining 17; 
our final answers do not depend on the undetermined fields. 

The $O(\epsilon)$ correction makes the D-terms nonzero, $D \sim \epsilon\, \phi^{(0)}\phi^{(1)}.$ Specifically, we find
\eqn\Dnew{
D_{U(1)} = 1.396\,{h^{8/5}   \Lambda^2\over g_1^2}  .
}
The D-terms for $SU(4)$ are also nonzero, but we will not record those here.
The corrected minimum of $V$ is
\eqn\Vnew{
V_{\rm min} = \left ( 2.22 
-  0.244 \, \epsilon_1 
 - 0.487 \, \epsilon_4 
 \right) h^{6/5} \Lambda^4 \,.
}

In addition to the  $U(1)_R$, the classical theory with
superpotential  
\Wdyn\ has a global $U(1)$
symmetry, 
which --- unlike the $U(1)_R$ --- is anomalous in the quantum theory. The $U(1)_R$ is broken in the vacuum while the global $U(1)$ is preserved. We have explicitly verified that the bosonic mass matrix has  
$15_{SU(4)}+1_{U(1)}+1_{U(1)_R}+1_{U(1)_{\rm global}}-(3_{SU(2)}+1_{U(1)_{\rm global}})=14$ zero modes. Of these Goldstone bosons, 13 are eaten by the Higgsed vectors, leaving $15-13=2$ remaining 
complex degrees of freedom. These matter degrees of freedom are singlets under the unbroken $SU(2)$. 

Let us summarize the dynamics as follows \PoppitzVD.
In the UV, we start with the 4-1 model. At the scale $\Lambda\, h^{-1/5}$, the matter fields get vevs and break the gauge group from $SU(4) \times U(1)$ to $SU(2)$. The Higgsed vectors acquire masses of order $g^2 \Lambda^2 h^{-2/5}$. The unbroken $SU(2)$ confines at the scale $\Lambda_2 = \Lambda \, h^{2/15}$ and the result is a low-energy sigma-model in which gaugino condensation produces the non-perturbative term in the superpotential. Finally, at the scale $E_{\rm susy} \sim \sqrt{F} \sim \Lambda\, h^{3/10}$, supersymmetry is broken. The hierarchy among these scales is guaranteed by $h \ll 1$.

\newsec{The Semi-Direct 4-1 Model}
\seclab\sdfourone

We now add to the 4-1 model $N_f$ pairs of messenger fields charged under the $U(1)$. We will denote the messenger fields by $L_\alpha$ and $\bar L^\alpha$, where $\alpha = 1, ..., N_f$. The messengers have charges $q$ and $-q$, respectively. We introduce (by hand) a mass term, so that the full superpotential is 
\eqn\wmass{
W = h \, B +2  \left ( {\Lambda^{10} \over Y} \right )^{1/2} + m L_\alpha \bar L^\alpha.
}
The mass term preserves a global $SU(N_f)$ symmetry which acts  
only on the messenger fields. 
The mass $m$ has to be large enough to ensure that the messengers do not acquire non-zero vevs. This requires 
\eqn\mconde{
  \epsilon_1~m_{V}^2~\,\roughly< ~\, m^2 \,\ll\, m_{V}^2\, .
}
where $m_{V}^2 \sim g^2 h^{-2/5} \Lambda^2$ is the mass squared of the Higgsed vector bosons. The smallness of $\epsilon_{1,4} = h^2/g_{1,4}^2$ is guaranteed by the assumption $h \ll g_1 \ll g_4 \ll 1$. The upper bound on $m$ is needed in order to keep the messengers in the low-energy theory obtained from integrating out the vectors. We derive the requirement \mconde\ in Section 5.

The messenger fields are coupled to the hidden sector only through gauge interactions, i.e.~only through the $U(1)$ D-terms. When SUSY is broken in the 4-1 model, the messenger masses are split and this is how SUSY breaking is communicated to the SSM. The purpose of this section is to calculate the messenger mass splittings. Following \SeibergQJ\ we do this in three different ways.

\subsec{Microscopic calculation}

The messenger masses get contributions from the nonzero $U(1)$ D-terms \Dnew\ calculated in the previous section. There are naively two possible types of mass terms: $m_{d}^2 L L^\dagger$ and $m_{od}^2 L \bar L$. We refer to these as ``diagonal" and ``off-diagonal" masses, respectively. Classically, only the diagonal masses are generated. They come from the $U(1)$ D-term via the cross-term
\eqn\mdmess{
g_1^2\, V^{U(1)}_{D} ~=~  {g_1^2 \over 8}   \left ( \sum_i q_i |\phi_i|^2 \right )^2 
~=~  {g_1^2 \over 4}  \, \langle D_{U(1)} \rangle \, q\, (|L_\alpha|^2 - |\bar L^\alpha|^2 )+ \dots
} 
Using \Dnew\ we find
\eqn\mUone{
 m_d^2  =0.349\, q \, \Lambda^2 \, h^{8/5} \, ,
}
so that messengers $L_\alpha$ and $\bar L^\alpha$ have diagonal masses $m^2 \pm m_d^2$.
Consistency requires that the messengers do not become tachyonic. This is ensured by $h$ being sufficiently small; the precise condition is given in Section 5.

It is worth noting that $m_d^2$ is independent of the gauge coupling.
That this must be so will be clear from the calculation in terms of gauge-invariant operators, which comes next.

\subsec{Macroscopic analysis with gauge-invariant operators}

We first re-analyze the pure 4-1 model in the language of gauge-invariant operators, and then add the messenger fields.

\subsubsec{The 4-1 Model}

In the pure 4-1 model, the independent $SU(4)$-invariants are $S$, $\,\bar F^i F_i$, and ${\rm Pf} A$. These have $U(1)$-charges 4, $-4$, and 4, respectively. 
Consequently, the only gauge invariants are $B = S \bar F^i F_i$ and $Y = {1 \over 4} \bar F^i F_i\, {\rm Pf} A$, and these parameterize the moduli space which we denote by ${\cal M}_0$. The point $Y=0$ is singular because the unbroken gauge group is larger and there are additional massless degrees of freedom. 
We are only interested here in points away from $Y=0$, so henceforth we assume $Y \ne 0$.
We can then form the real dimensionless operator 
\eqn\theT{
T = |Y|^{-3/2} \, B^\dagger B, 
}
and by dimensional analysis, the classical K\"ahler potential must be of the form 
\eqn\KzeroA{
 K_{{\cal M}_0} = |Y|^{1/2} K_0 (T) \,. 
}

The function $K_0(T)$ is determined as follows. 
We can go to a point on moduli space where the $SU(4)$ and $U(1)$ D-flatness conditions let us write $B = b^2 c$ and $Y =a^2 b^2$, with $c^2 = b^2 - a^2/2$. 
We can then set 
\eqn\KzeroB{
  a^2 = Y^{1/2} f(T)\, ,~~~~~~~
  b^2 = \frac{Y^{1/2}}{f(T)}  \, .
}
Note that $0< a^2 \le 2 b^2$ requires $0<f(T) \le \sqrt{2}$.

Comparing \KzeroB\ with $T = (b^4 c^2)/(a^3 b^3)$, we see that $f$ satisfies
\eqn\KzeroB{
  T f^3 + \frac{1}{2} f^2 - 1 = 0 \, .
}
The real solution of \KzeroB\ is
\eqn\funcf{
  f(T) = \frac{1}{6T}
  (-1 + h_-(T)^{1/3} + h_+(T)^{1/3}),
}
with $h_\pm(T) = -1 + 6T \big[18 T \pm \sqrt{324 T^2 - 6}\,\big]$. This solution satisfies the necessary bound on $f(T)$.

As explained at the end of Section 2, the low-energy theory is described by a non-linear sigma model with no remaining gauge degrees of freedom. In Wess-Zumino gauge, the matter kinetic terms are simply the canonical terms, so the K\"ahler potential is$\,$\foot{The factor $1/2$ for the antisymmetric field comes from the normalization of the K\"ahler term, 
$-\frac{1}{2}(A^\dagger)^{ij} (e^V)_{ij}^{~~kl} A_{kl}$. It ensures canonical normalization of the independent components of $A$.}

\eqn\lagr{
  |F|^2 + |\bar F|^2 + \frac{1}{2} |A|^2 + |S|^2
  ~=~ 3b^2 + \frac{1}{2} a^2
  ~=~ |Y|^{1/2} \left[ \frac{3}{f(T)} + \frac{f(T)}{2} \right]
}
where $|F|^2 \equiv F_i {F^\dagger}^i$ and $\,|A|^2 \equiv {\rm Tr}\,A^\dagger A$.

Comparing \KzeroA\ and \lagr, we can read off
\eqn\KzeroC{
   K_0(T) =
   \frac{3}{f(T)} + \frac{f(T)}{2}
    \, .
}

The scalar potential is 
\eqn\WgiA{
  V_0 = g_0^{A \bar B}\,\partial_A W\, \partial_{\bar B} \overline W \, ,
}
where $g_0^{A \bar B}$ is the inverse of the K\"ahler metric
${g_0}_{A \bar B} = \partial_A\, \partial_{\bar B}\, K_{{\cal M}_0}$. Evaluating $V_0$ at the values of $a$ and $b$ given in \minA\ we find $V_{0, \rm min} = 2.22\,  h^{6/5} \Lambda^4$, 
in agreement with the microscopic calculation.


\subsubsec{$U(1)$-charged messenger fields}

Let us now add to the 4-1 model $N_f$ pairs of $SU(4)$-singlet messenger fields $L_\alpha$ and $\bar L^\alpha$ with $U(1)$ charges $\pm q$. We will make the convenient choice $q = 4$.
The gauge-invariant operators are then
\eqn\GIop{
  Y = \frac{1}{4}\, \bar F^i F_i \,{\rm Pf} A\, ,
  ~~~~ X_a = {\cal L}_a \, \bar F^i F_i \,, 
  ~~~~ Z^\alpha = {\bar L}^\alpha \,{\rm Pf} A\, ,
  ~~~~ R_a^{~\alpha} = {\cal L}_a {\bar L}^\alpha \, .
}    
Here $\alpha = 1, \dots , N_f$ and $a = 1, \dots, N_f+1$. We have introduced the notation ${\cal L}_a = L_a$ for $a = 1, \dots, N_f$ and ${\cal L}_{a=N_f+1} = S$. These fields are related by the classical constraint 
\eqn\constraintA{
X_a Z^\alpha - 4\, Y R_a^{~\alpha} = 0\, . 
}
Assuming as above that $Y\ne 0$, we can solve the constraint \constraintA\ and eliminate $R_a^{~\alpha}$. Thus our independent gauge-invariant operators are $Y$, $X_a$ and $Z^\alpha$. These parameterize the moduli space ${\cal M}$ of the model with messengers.

The moduli space ${\cal M}_0$ discussed in the previous section is the subspace of ${\cal M}$ corresponding to setting $Z^\alpha = X_\alpha = 0$ for $\alpha = 1, \dots , N_f$. To obtain the messenger masses, we can expand around ${\cal M}_0$ and find the K\"ahler metric on ${\cal M}$ in the neighborhood of ${\cal M}_0$. For the purpose of making contact with the results in the previous section, it is convenient to identify $X_{N_f+1} = B$ and work with the dimensionless variable $T$ defined in \theT.
Near ${\cal M}_0$, the K\"ahler potential takes the general form
\eqn\KpotA{
  K_{\cal M} =  |Y|^{1/2} K_0(T) 
  + |Y|^{-1} \Big[ K_1(T) X^{\dagger\alpha} X_\alpha
  + K_2(T) Z^\alpha Z^{\dagger}_\alpha 
  + K_3(T) (Z^\alpha X_\alpha + (Z^\alpha X_\alpha)^\dagger) \Big].
}
There are corrections to this starting at quartic order in $X$ and $Z$.
$K_0$ is given in $\KzeroC$.

To leading order, $X_\alpha = b^2 L_\alpha$ and 
$Z^\alpha = 4 a^2 {\bar L}^\alpha$. From the canonical kinetic terms $|L|^2 + |\bar L|^2$ we can read off the functions $K_{1,2,3}$, which are 
\eqn\KpotB{
  K_1(T) = f(T)^2 \, ,
  ~~~~~~~ 
  K_2(T) = \frac{1}{16 f(T)^2},
  ~~~~~~~
  K_3(T) = 0\, .
}

It is convenient to rescale the fields to separate out the $Y$-dependence. Thus we define
\eqn\resc{
  {\hat X}_\alpha = Y^{-1/2} X_\alpha \, ,~~~~~~~
  {\hat Z}^\alpha = Y^{-1/2} Z^\alpha \, ,~~~~~~~
  {\hat B} = Y^{-3/4} B \, .
}
In these variables,  $T={\hat B}^\dagger {\hat B}$ and the K\"ahler potential takes the simple form
\eqn\KpotC{
  K_{\cal M} =  |Y|^{1/2} K_0(T) 
  + K_1(T) {\hat X}^{\dagger\alpha} {\hat X}_\alpha
  + K_2(T) {\hat Z}^\alpha {\hat Z}^{\dagger}_\alpha 
  + \dots
}
where ``\dots" represents higher order corrections.

Let us now consider the superpotential \wmass\ with the mass term $m \bar L^\alpha L_\alpha = Z^\alpha X_\alpha/(4 Y)$. After the rescalings \resc, the full superpotential is
\eqn\WgiB{
 W = h\, Y^{3/4}\, \hat B + 2 \Lambda^5\, Y^{-1/2} 
   + \frac{m}{4} \, \hat{Z}^\alpha \, \hat X_\alpha  
 \, .
}

Finally we can calculate the scalar potential $V$ expanded to quadratic order in the messengers $X_\alpha$ and $Z^\alpha$. The mass terms can be read off from the potential, but one must take into account the non-canonical kinetic terms arising from the K\"ahler potential. Cross-terms do not arise because $K_3=0$.
We find that SUSY breaking produces only diagonal mass terms for the messengers. Specifically,  $m_X^2 = m^2 + m_d^2$ and $m_Z^2 = m^2 - m_d^2$, where
\eqn\msgmass{
 m_d^2 = - \frac{4 f(T) \big[f(T)^2 -6 \big]^2 
 \big[ 2\, T f(T) \Lambda^5 + \sqrt{T} \,Y^{5/4} h\big]^2}
 {Y^2\, T f'(T) \big[ 6+ 36 T f(T) + f(T)^2 \big]^2} \, .
}
Evaluating \msgmass\ on the solution \minA, we find 
$m_d^2 = 1.396\,  h^{8/5} \Lambda^2$. This agrees with the result \mUone\ of the microscopic calculation when $q = 4$.

In this calculation of $m_d^2$, the gauge fields were integrated out, ignoring\foot{This is equivalent to taking $\epsilon \to 0$ in the microscopic calculation.} their kinetic terms $\frac{1}{2g^2} W_\alpha W^\alpha$. Therefore the result cannot depend on the gauge couplings; this explains why $g_1^2$ had to drop out of the microscopic calculation of $m_d^2$. 
Note that $m_d^2$ arises  from the curvature on moduli space, while in the microscopic calculation it comes from the non-vanishing of the D-terms in Wess-Zumino gauge.


\subsec{Unitary gauge}

In our third calculation of the messenger masses we find the effective K\"{a}hler potential that comes from integrating out the massive vectors in unitary gauge. (For an early reference on 
effective K\"{a}hler potentials in unitary gauge, see \PoppitzTX.) We denote the  $U(1)$ vector superfield by $U$ and that of $SU(4)$ by $V=V_a T^a$. The K\"ahler terms in the pure 4-1 model are
\eqn\lagr{
  S^\dagger e^{q_S U} S
  + (F^\dagger)^i e^{q_F U} (e^{V_a T^a_{\sfund}})_i^{~j} F_j
  + (\bar F^\dagger)_i e^{q_{\bar F} U} 
  (e^{V_a T^a_{\overline{{\sfund}}}})^i_{~j} \bar F^j
  - \frac{1}{2} (A^\dagger)^{ij} 
  e^{q_{A} U} (e^{V_a T^a_{\sasym}})_{ij}^{~~kl}
  A_{kl}.
}

In the limit $g_{1,4} \to \infty$ we can ignore the kinetic terms of the gauge fields, so the equations of motion for the gauge fields arise only from \lagr.
 It is convenient to denote the gauge fields corresponding to the broken generators\foot{The splitting of the 15 $SU(4)$ generators into 12 broken and 3 unbroken generators is given explicitly in Appendix B.} by $V_I$, where $I=1,...,13$, with
$V_I = V_a$ for $I=1,\dots,12$ and $V_{13} = U$. 
We then write the equations of motion for $V_I$ as
\eqn\eomVI{
  0 = \hat{D}^I + \lambda^{IJ} V_J  + \dots
  ~~~~\to~~~~
  V_I = - \lambda^{-1}_{IJ}  \,\hat{D}^J + \dots\, ,
}
where ``+ \dots" denotes higher order terms. In this equation,
$\hat{D}^J$ are the D-terms,
\eqn\hatDs{
 \eqalign{
  \hat D^a &=   
  (F^\dagger)^i (T^a_{\sfund})_i^{~j} F_j
  - \bar F^i (T^a_{\sfund})_i^{~j} \bar F^\dagger_j
  - \frac{1}{2} (A^\dagger)^{ij} 
  ({T^a_{\sasym}})_{ij}^{~~kl}
  A_{kl} \, ,\cr
  \hat D^{13}&=
    q_S\, S^\dagger  S
  + q_F\, (F^\dagger)^i  F_i
  + q_{\bar F} \, \bar F^i \bar F^\dagger_i
  - \frac{1}{2} q_A\,  (A^\dagger)^{ij} A_{ij},
}
}
and the vector mass matrix $\lambda^{IJ}$ has the following components:
\eqn\lamab{
\eqalign{
  \lambda^{13,13} &=
      q_S^2\, S^\dagger  S
  + q_F^2\, (F^\dagger)^i  F_i
  + q_{\bar F}^2 \, \bar F^i \bar F^\dagger_i
  - \frac{1}{2} q_A^2 \,  (A^\dagger)^{ij} A_{ij}\, ,\cr
   \lambda^{13,a} &=
   q_F \, (F^\dagger)^i   (T^a_{\sfund})_i^{~j} F_j
  - q_{\bar F} \, \bar F^i\, (T^a_{\sfund})_i^{~j} \bar F^\dagger_j
  - \frac{1}{2} q_A \,  (A^\dagger)^{ij}  ({T^a_{\sasym}})_{ij}^{~~kl} A_{kl}\, ,\cr
   \lambda^{ab} &=
   \frac{1}{2}  (F^\dagger)^i   
   \{ T^a_{\sfund}, T^b_{\sfund}\}_i^{~j} F_j
  + \frac{1}{2} \, \bar F_i\, \{ T^a_{\sfund}, T^b_{\sfund}\}_i^{~j} \bar F^\dagger_j
  - \frac{1}{4} \,  (A^\dagger)^{ij}   
    \{ T^a_{\sasym}, T^b_{\sasym} \}_{ij}^{~~kl} A_{kl}\, .
}}
By excluding the generators of the unbroken $SU(2)$ subgroup, we are ensuring that $\lambda^{IJ}$ is invertible.
Note also that because the $\hat{D}^I$ are functions of $\Phi^\dagger$ and $\Phi$, we will occasionally write $\hat{D}^I=\hat{D}^I(\Phi^\dagger,\Phi)$. 

It is useful to consider the superfield equation \eomVI\ in component form. Reading off the $\theta^2 \bar \theta^2$ component, we see that $D \sim |{\cal F}|^2/|\phi|^2$, where ${\cal F}$ is shorthand for the F-term components of the chiral field $\Phi$ and $\phi$ stands for the lowest component of $\Phi$. This makes it transparent that the non-vanishing D-terms are induced by the SUSY-breaking F-terms. Additionally, we see that the $\theta=\bar \theta=0$ component of the vector is nonzero. This is why the physics of a massive vector superfield is clearer in unitary gauge than in Wess-Zumino gauge.

Substituting into the Lagrangian gives the effective K\"ahler potential
\eqn\KeffA{
 K_{\rm eff} = K^{(0)} - \frac{1}{2} \hat{D}^I \lambda^{-1}_{IJ} \hat{D}^J + \dots
}
with canonical contribution $K^{(0)} = S^\dagger  S
  + (F^\dagger)^i  F_i
  + \bar F^i \bar F^\dagger_i
  - \frac{1}{2} (A^\dagger)^{ij} A_{ij}$.

Including messengers $L_\alpha$ and $\bar L_\alpha$, we now expand around the D-flat directions $\phi_0$. In our notation, $\hat D^I (\phi_0^\dagger, \phi_0) = 0$. Writing a general field as $\Phi = \phi_0 + \delta \Phi$, we impose the unitary gauge condition\foot{This condition would be trivially satisfied for the unbroken generators and thus would impose no constraints on the fluctuations $\delta \Phi$.} $\phi_0^\dagger T^I \delta \Phi =0$. Equivalently, we can write  
$\hat D^I (\phi_0^\dagger,\delta \Phi) =0$.  In this gauge, the effective K\"ahler potential is
\eqn\KeffB{
 \eqalign{
 K_{\rm eff} &= K^{(0)} 
 + L^{\dagger\alpha} L_\alpha + \bar{L}_\alpha^\dagger\bar{L}^\alpha 
 - \frac{1}{2} \delta \hat{D}^I \lambda^{-1}_{(0)IJ} \,\delta \hat{D}^J 
 - q\, ( L^{\dagger\alpha} L_\alpha 
 - \bar{L}_\alpha^\dagger\bar{L}^\alpha )\,
 \lambda^{-1}_{(0)13,J} \,\delta \hat{D}^J 
 + \dots
}
}
with $\delta \hat{D}^I \equiv \hat{D}^I (\delta\Phi^\dagger,\delta\Phi)$ and $\lambda_{(0)}$ is \lamab\ evaluated at $\phi_0$.

We obtain the K\"ahler metric $g_{A\bar B}$ from $K_{\rm eff}$ by differentiating with respect to the fluctuations $\delta \Phi$ and $L_\alpha, \bar L^\alpha$. The masses of the messengers are then obtained from the effective potential
\eqn\VeffUG{
V_{\rm eff} = g^{A\bar B} \partial_A W \, \partial_{\bar B}  \bar W \, .
}
The leading order terms in $K_{\rm eff}$ are canonical, so it is trivial that the vacuum energy $V_{\rm min}$ agrees with the microscopic calculation. 
The SUSY-split messenger masses arise from the final term in \KeffB\ via the nonzero D-term. Specifically, \VeffUG\ gives diagonal masses of the form
\eqn\msgUG{
m_d^2 ~=~ 
\frac{2b^2-a^2+a^3 b^3 \left(a b^5+2c-a b^3 c^2\right)}
{4 a^4 b^4 \
\left(a^2+2 b^2+4 c^2\right)} \, q 
~=~ {4 b^2-2 a^2  +a^6 b^6 + 2 a^3 b^3 \sqrt{4 b^2 - 2 a^2} \over 2 a^4 b^4 (6 b^2-a^2)} \, q .
}
Evaluating $a$ and $b$ at the minimum of the potential, we find that this agrees with the previous calculations. It is clear from the form of the K\"ahler potential  \KeffB\ that no off-diagonal terms are generated.

\subsec{Radiative Corrections}
Including one-loop corrections to the K\"{a}hler potential is straightforward. 
This will produce off-diagonal masses and a nonzero supertrace over the messenger sector.

We can set $g_4=0$ for the purpose of computing the leading order radiative corrections to the messenger masses. 
The one-loop correction \refs{\GrisaruVE,\IntriligatorDD} relevant for the messenger masses is 
\eqn\kloopGris{
K^{\rm 1-loop}=
\frac{1}{(4\pi)^2} \,2 \, g_1^2\, \tr\, M^2 \log (M^2/\Lambda_1^2) \, ,~~~~~~~
M^2 =  \sum_i q_i^2  |\phi_i|^2 \, ,
}
where the sum is over the $U(1)$-charged spectrum.\foot{Since SUSY is broken, there are corrections to this formula, but they are suppressed by $h \ll 1$.} 
For the 4-1 model with $N_f$ pairs of messenger fields we find
\eqn\kloop{
M^2 =  16 |S|^2 + 9 |F|^2 + |\bar F|^2 + 2 |A|^2 
+ q^2 
( | L_\alpha |^2 + | \bar  L^\alpha |^2 ) \, .
}
The sum over $\alpha = 1,\dots, N_f$ is implicit.

Defining $\alpha_1\equiv g_1^2/4\pi$ and expanding the K\"ahler potential to quadratic order in the messenger fields, we find
\eqn\kloop{
K^{\rm 1-loop} = K_0^{\rm 1-loop} 
+ {\alpha_1 \over 2 \pi} \, q^2 \left ( |L_\alpha|^2 + |\bar L^\alpha |^2 \right ) 
\left( 1 + 
\log\left[ { 16 |S|^2 + 9 |F|^2 + |\bar F|^2 + 2 |A|^2 \over \Lambda_1^2 }\right]\right) \, .
}
The first term, $K_0^{\rm 1-loop}$, is independent of the messenger fields and contains the one-loop correction to $K_0$ in \KzeroC.
We are interested here in two quantities, the off-diagonal messenger masses and the supertrace of the messenger sector. These two quantities are both unaffected by $K_0^{\rm 1-loop}$, which we will therefore not concern ourselves with any further.

The second term of \kloop\ corrects $K_{1,2}$ of the K\"ahler potential \KpotA. In terms of the gauge-invariant operators, we can write
\eqn\konetwo{
  K_{1,2}^{\rm 1-loop}
  = {\alpha_1 \over 2 \pi}\, q^2 \, K^{\rm cl}_{1,2} \,
  \bigg[ 1+ 
  \log \bigg( \frac{ 2 |Y|^{1/2} \big[13 - 2 f(T)^2 \big]}{f(T) \, \Lambda_1^2} \bigg) \bigg],
}
where we must take $q=4$ for consistency with our analysis in Section 3.1.

The off-diagonal messenger masses $m^2_{\rm od} X_\alpha Z^\alpha$ are now non-vanishing,
\eqn\mod{
  m^2_{\rm od} ~=~ \frac{\alpha_1}{\pi} \,
  \frac{(9 \Lambda^5 - 13 h \,\sqrt{T} \, |Y|^{5/4}) f(T)}
  {|Y| (13 - 2 f(T)^2)}
  \, m\, q^2 
  ~=~
  0.0128\, q^2 \, \alpha_1\,  h^{4/5}\, m \, \Lambda \, .
}

The diagonal masses $m_d^2$ also receive contributions from the corrected K\"ahler potential. This leads to a negative supertrace over the messenger sector. Specifically,
\eqn\Str{
  {\rm Str}\, m_{\rm msg}^2 = \tr\, M_0^2 - \tr\, M_{1/2}^2 
  = -0.374 \, \alpha_1 \, q^2 \, N_f \, h^{8/5} \Lambda^2 \, .
}



\newsec{Adding a Fayet-Illiopoulos term}

We now consider the 4-1 model with an FI term $\xi$. Within the considered range of parameters, the vacuum breaks SUSY for all values of $\xi$. 
In this section, we derive the SUSY-split messenger masses.


\subsec{Microscopic calculation with FI term}

The addition of the FI term changes the $U(1)$ D-flatness condition to be
\eqn\DUxi{
  2 a^2 - 4b^2 + 4 c^2 + \xi = 0 \, .
}
Scaling all fields as above \VtotA , the scalar potential $V$ now depends on a new dimensionless quantity 
$\xi'=\xi\, h^{2/5}\Lambda^{-2}$. Assuming $\epsilon = h^2/g^2 \ll 1$, the minimum of $V$ is located near the D-flat directions. When $\xi'$ is large, we must also assume (see Section 5) 
\eqn\xim{
\frac{\xi'}{8} 
<  \frac{1}{h^{8/5}}\frac{m^2}{\Lambda^2}  
\ll  \frac{1}{\epsilon}\, \frac{1}{\xi'^{2/3}} \qquad {\rm for\,\, large}~\xi^\prime .
}
in order for the messengers to have zero vevs.
Note that a necessary condition is $\epsilon \ll \xi'^{-5/3}$.

\bigskip
\centerline{\epsfxsize=0.45\hsize\epsfbox{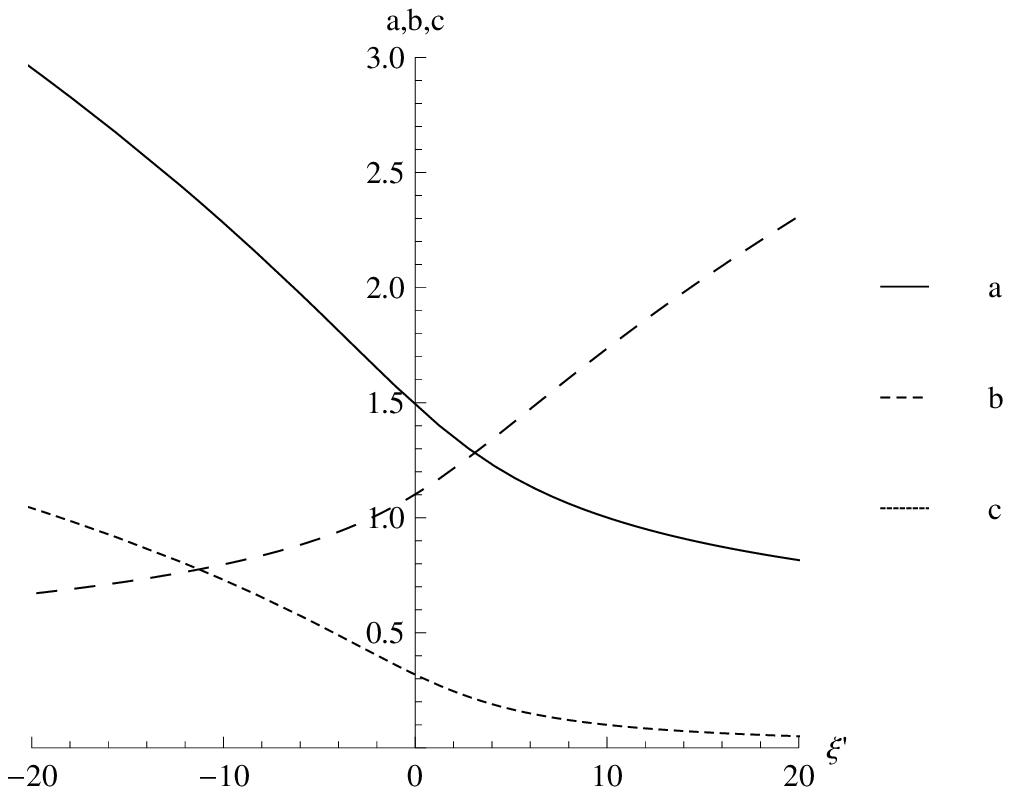}
\epsfxsize=0.45\hsize\epsfbox{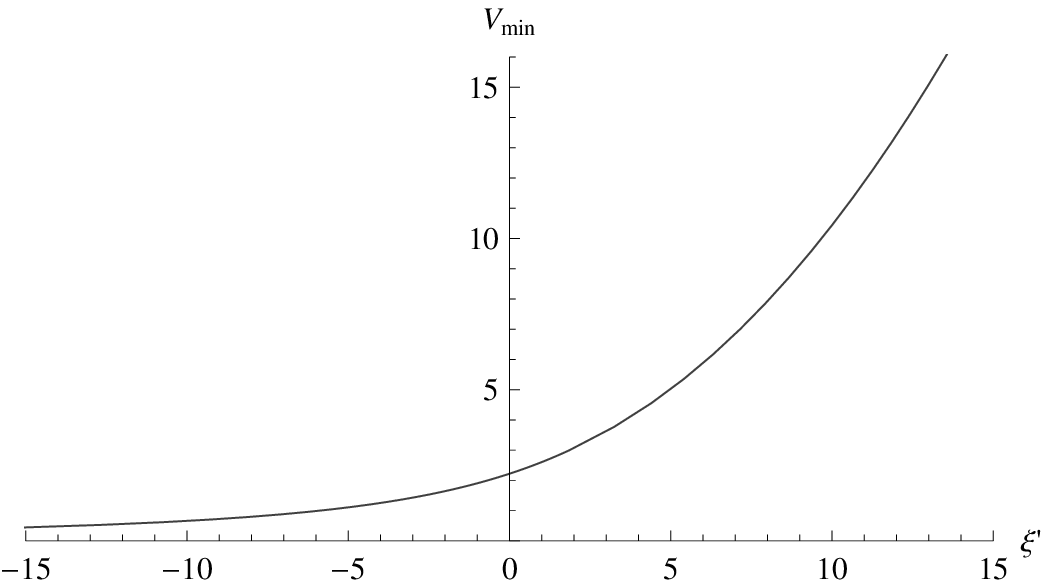}
}
{\ninepoint\sl \baselineskip=2pt  {\bf Figure 1 (left):}
{\sl Solutions of $a$, $b$ and $c$ at the minimum of the potential vs.~$\xi'
\equiv \xi\, \Lambda^{-2} h^{5/2}$.}}

{\ninepoint\sl \baselineskip=2pt  {\bf Figure 1 (right):}
{\sl $V_{\rm min}$ vs.~$\xi'$.}}
\bigskip

\bigskip
\centerline{\epsfxsize=0.45\hsize\epsfbox{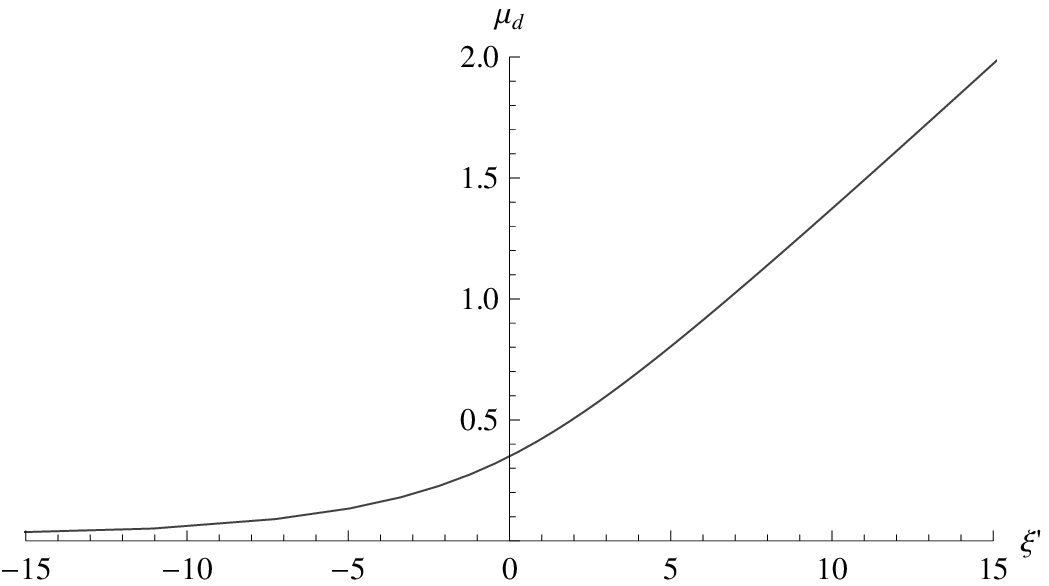}
}
\centerline{\ninepoint\sl \baselineskip=2pt {\bf Figure 2:}
{\sl Diagonal  messenger mass splittings $\mu_d(\xi')$.}}
\bigskip

To leading order we minimize the F-term potential on the D-flat directions. The solutions for $a$, $b$ and $c$ at the minimum are displayed as functions of $\xi'$ in Figure 1. The value of $V_{\rm min}$ (in units of $h^{6/5} \Lambda^4$) is a monotonically increasing function of $\xi'$; see Figure 1. 

The $O(\epsilon)$-corrections to the potential are calculated as without the FI-term. As before, the non-vanishing of the $U(1)$ D-term at the corrected minimum of the potential introduces SUSY-breaking mass splittings for the messengers: The diagonal mass terms are $m^2 \pm \mu_d(\xi')\, |q| \, h^{8/5}\Lambda^2$. 
The function $\mu_d(\xi')$ grows monotonically with $\xi'$, see Figure 2. 
Consistency requires that the messengers do not become tachyonic, so there is a bound on how large $\xi$ can be.  We will return to this in 
Section 5.2.


\subsec{Unitary gauge}

The calculation in unitary gauge
proceeds as in Section 3.3 and we will only highlight the changes resulting from having $\xi \ne 0$.
The K\"ahler terms \lagr\ are modified by the additional FI-term $+\xi U$. The equations of motion for the gauge fields again take the form \eomVI\ with the matrix $\lambda^{IJ}$ unchanged, i.e.~given by \lamab, and  only the $U(1)$ D-term modified, 
$\hat{D}_\xi^a = \hat{D}^a$  and $\hat{D}_\xi^{13} = \hat{D}^{13} + \xi$. The unitary gauge condition for the fluctuations $\delta \Phi$ around the D-flat vacuum $\phi_0$ now reads
$\hat{D}_\xi^I(\phi_0^\dagger,\delta \Phi) =0$.
When the messenger fields with vanishing vevs are included, the effective K\"ahler potential is 
\eqn\KeffBxi{
 \eqalign{
 K_{\rm eff} &= K^{(0)} 
 + L^{\dagger \alpha} L_\alpha + \bar{L}_\alpha^\dagger\bar{L}^\alpha 
 - \frac{1}{2} \delta \hat{D}_\xi^I \lambda^{-1}_{(0)IJ} \,\delta \hat{D}_\xi^J 
 - q\, ( L^{\dagger \alpha} L_\alpha - \bar{L}_\alpha^\dagger\bar{L}^\alpha  )\,
 \lambda^{-1}_{(0)13,J} \,\delta \hat{D}_\xi^J 
 \cr
 & \quad\quad\quad
 + \xi\, q\, \lambda^{-1}_{(0)13,13} \, (L^{\dagger \alpha} L_\alpha - \bar{L}_\alpha^\dagger\bar{L}^\alpha)
 + \dots
}
}
with $\delta \hat{D}^I \equiv \hat{D}^I (\delta\Phi^\dagger,\delta\Phi)$. Compare this result with \KeffB\ to note the new $\xi\, q$ contribution. The result for the messenger mass splittings is
\eqn\msgUGb{
m_d^2 = 
\frac{4 b^2-2 a^2+2 a^3 b^3 \sqrt{4 b^2-2 a^2-\xi }+a^4 b^6 
\left(a^2+\xi/2 \right)}{2 a^4 b^4 \left(6 b^2-a^2-\xi \right)} \,\, q 
\, .
}
which agrees numerically with the result found in the microscopic calculation.


\subsec{Gauge-invariant operators}

We now wish to reproduce \msgUGb\ via gauge-invariant operators. 
To begin, we write the K\"{a}hler potential compactly as 
\eqn\kisii{
\eqalign{
K & =S^\dagger e^{4U} S +  \tr \Big\{ F^{\dagger} e^{-3U} e^{V^A T_{\sfund}^A} F + \bar F e^{-U} e^{-V^A T_{\sfund}^A} \bar F^\dagger + {1 \over 2} A^\dagger e^{2U} e^{V^A T^A_{\sasym}} A \Big\}+ \xi U  \cr
&\equiv K_S + K_F + K_{\bar F} + K_A + \xi U \, .
}}
The FI-term $\xi U$ makes the K\"{a}hler potential \kisii\ gauge-dependent.
Instead of working in Wess-Zumino gauge, as we did in Section 3.2, we will here avoid an explicit gauge choice and proceed by a different method\foot{We are grateful to N.~Seiberg for showing us this technique.} to integrate out the gauge fields to obtain the K\"ahler potential in terms of the gauge-invariant operators $Y$ and $B$.

We need to express $U$ and $K_{S,F,\bar F,A}$  in terms of $\xi$,  $|B|^2=B^\dagger B$ and $|Y|^2=Y^\dagger Y$. The $SU(4)$ D-flatness conditions imply that $\bar F^i e^{-U} \bar F^\dagger_j = F^{\dagger i}e^{-3U} F_j$
and $(A^{\dagger})^{ik} e^{2U} A_{jk} \propto \delta^i_j$. The former gives $ K_{\bar F} = K_F$ and the latter $|{\rm Pf}\, A|^2 = 4 ({\rm Tr} A^\dagger A)^2$. These relations are needed to show that $|Y|^2 = K_F^2 K_A^2$ and $|B|^2 = K_S K_F^2$.

The $U(1)$-flatness condition, $4 K_S - 3 K_F - K_{\bar F} + 2 K_A + \xi =0$, is simply the equation of motion for the $U(1)$ gauge field, ${\partial K/\partial U} = 0$, in the limit where we neglect the gauge kinetic terms. We use the above results to write the $U(1)$-flatness condition as a cubic equation which determines $K_F$ in terms of $|B|$, $|Y|$, and $\xi$:
\eqn\keqnii{ K_F^3 - {\xi \over 4} K_F^2 -  {1 \over 2} |Y| K_F -  |B|^2 =0.
}
Let us introduce $T \equiv |B|^2 / |Y|^{3/2}$ and $y \equiv \xi / (4 |Y|^{1/2})$, and set $f(T,y) \equiv |Y|^{1/2}/K_F$. Then \keqnii\ 
becomes
\eqn\keqnres{
T f^3 + \frac{1}{2} f^2 + y f - 1 = 0 \, .
}
Note that for vanishing FI term, \keqnres\ reduces to equation 
\KzeroB\ for $f(T)=f(T,y=0)$ (see Section 3.1).  
When solving \keqnres , we must choose the real positive root.

To finish the calculation of the K\"{a}hler potential, we need to solve for $U$. Note that $\log K_S = 4U + \log S + \log S^\dagger$. Using $K_S = |B|^2/K_F^2 =  f(T,y)^2\,  |B|^2/ |Y|$ we have
$U = {1 \over 2} \log f(T,y)$ + {hol.} + {anti-hol.} This contributes to the K\"ahler potential only through the term $\xi U$, so the purely holomorphic and anti-holomorphic terms can be dropped. Thus we have obtained the K\"ahler potential in terms of the gauge-invariant operators, 
\eqn\kult{
K_0^\xi ~=~  K_S + K_F + K_{\bar F} + K_A + \xi U 
~=~ |Y|^{1/2} \bigg( \frac{3}{f(T,y)}  + \frac{f(T,y)}{2} \bigg)
+ {1 \over 2} \xi \log  f(T,y) \, .
}
In the second equality, we have dropped a constant term $|Y|^{1/2} y = \xi/4$. It is clear that when $\xi=0$, the K\"ahler potential  
\kult\ reduces to \KzeroA, \KzeroC .
A useful test of the correctness of $K_0^\xi$ is that it produces the correct minimum value of the scalar potential $V=g^{A\bar B} \partial_A W \partial_{\bar B} \overline{W}$ with the K\"ahler metric $g_{A\bar B}$ obtained from \kult\ and the superpotential $W = h B + 2{\Lambda^5 \,Y^{-1/2}}$. Our result \kult\ has passed this qualifier exam.

\subsubsec{And now with messengers}

Adding the messenger fields with $U(1)$ charges $\pm 4$ to the 4-1 model gives rise to the gauge-invariant operators described in \GIop\ and \constraintA . We include as before the superpotential mass term
$W_m = m L_\alpha \bar L^\alpha = m\, X_\alpha Z^\alpha/(4Y)$.

The Lagrangian contains kinetic terms for the messengers $K_L + K_{\bar L}$, where $K_L = L^{\dagger\alpha} e^{4U} L_\alpha$ and
$K_{\bar L} = {\bar L^\dagger}_\alpha e^{-4U} \bar L^\alpha$.
$K_L$ and $K_{\bar L}$ satisfy
\eqn\XXZZ{
   |Y|^2 K_L K_{\bar L} = \frac{1}{16} (X^{\dagger\alpha} X_\alpha) (Z^{\dagger}_\beta Z^\beta) \, ,~~~~~~~~~
   K_L K_F^2  = X^{\dagger\alpha} X_\alpha
   \, .
}
It follows that
\eqn\KLs{
  K_L = K_F^{-2}  \, X^{\dagger\alpha} X_\alpha\, ,
  ~~~~~
  K_{\bar L} = \frac{K_F^2}{16 |Y|^2} \, Z^{\dagger}_\beta Z^\beta.
}
These results hold to leading order in the neighborhood of the D-flat directions of the pure 4-1 model. In this neighborhood, the K\"ahler potential is
\eqn\fullK{
  K^\xi = K_0^\xi
   + K_F^{-2}  \, X^{\dagger\alpha} X_\alpha
   + \frac{K_F^2}{16 |Y|^2} \, Z^{\dagger}_\beta Z^\beta + \dots
}
where ``\dots'' denotes higher order terms in $X$ and $Z$.
Here $K_0^\xi$ is given in \kult\ and $K_F = f(T,y)^{-1} |Y|^{1/2}$.

The K\"ahler metric $g_{A\bar{B}}$ is a $(2+N_f) \times (2+N_f)$ matrix. From its inverse we compute the scalar potential, and in particular the mass terms of the messenger fields. The result confirms the two other calculations.


\subsec{Radiative corrections}

It is straightforward to calculate of the one-loop correction to the
K\"ahler potential when $\xi \ne 0$.
The second term of \kloop\ corrects $K_{1,2}$ of the K\"ahler potential. In terms of the gauge-invariant operators we can write it
\eqn\konetwo{
  K_{1,2}^{\rm 1-loop}
  = {\alpha_1 \over 2 \pi}\, q^2 \, K^{\rm cl}_{1,2} \,
  \bigg[1+
  \log \bigg( \frac{ 2 |Y|^{1/2} \big[13 - 2 f(T,y)^2 - 8 y f(T,y) \big]}{f(T,y) \, \Lambda_1^2} \bigg)
 \bigg],
}
with $q = 4$.
The off-diagonal masses $m_{\rm od}^2$ and the supertrace ${\rm Str}\, m^2_{\rm msg}$ of the messenger sector can then be calculated; the results depend on $\xi$ through $y=\xi/(4|Y|^{1/2})$.

\bigskip
\centerline{\epsfxsize=0.43\hsize\epsfbox{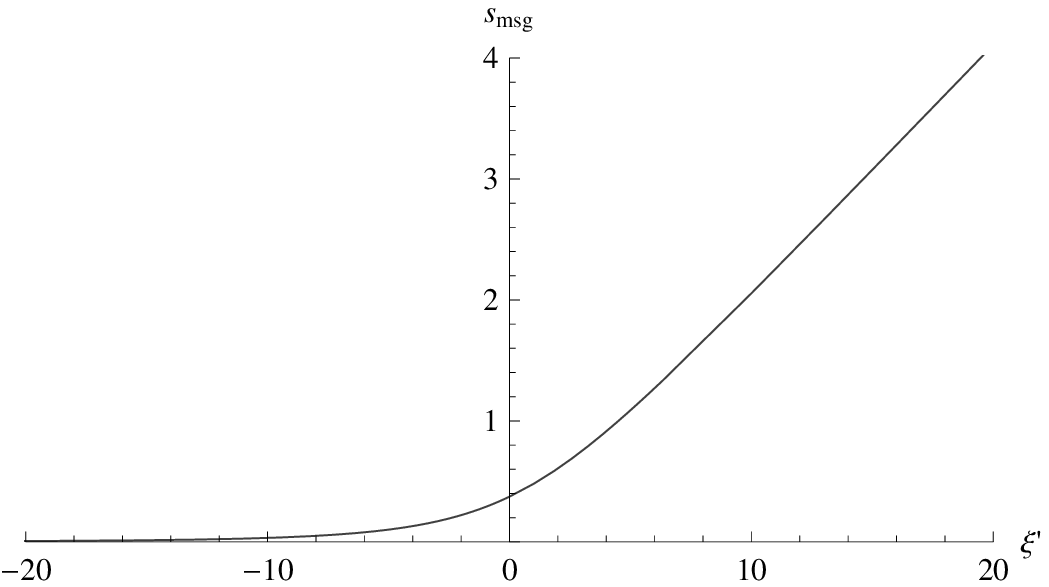}~~~~
\epsfxsize=0.48\hsize\epsfbox{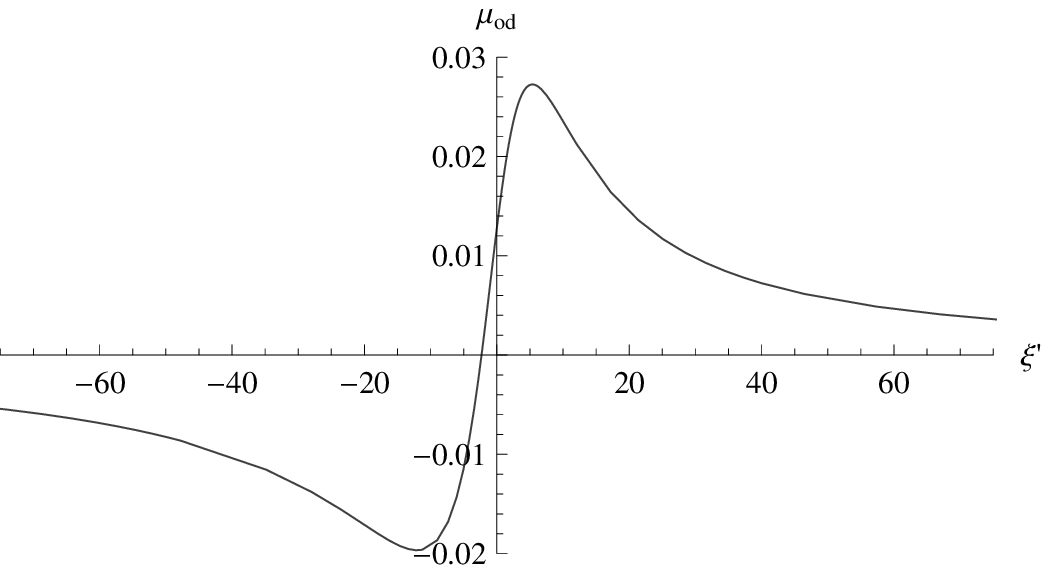}
}
\noindent {\ninepoint\sl \baselineskip=2pt {\bf Figure 3 (left):} 
{\sl Supertrace of the messenger sector $s_{\rm msg} \equiv (-{\rm Str}\, m^2_{\rm msg})/(\alpha_1 q^2 N_f h^{8/5} \Lambda^2)$ vs.~$\xi'$.}}

\noindent {\ninepoint\sl \baselineskip=2pt {\bf Figure 3 (right):}
{\sl Off-diagonal messenger masses $\mu_{\rm od} \equiv m_{\rm od}^2/(q^2 \alpha_1 h^{4/5} m \Lambda)$ vs.~$\xi'$. }}
\bigskip

We display the behavior of $m_{\rm od}^2$ and ${\rm Str}\, m^2_{\rm msg}$ in Figure 3. 
The supertrace is negative for all values of $\xi'=\xi\, \Lambda^{-2} h^{2/5}$; it approaches zero as $\xi' \to 0$ and decreases monotonically as $\xi' \to \infty$.
The off-diagonal masses have a more interesting behavior.
As is clear from Figure 3, $m_{\rm od}^2 =\mu_{\rm od} \,q^2 \alpha_1 h^{4/5} m \Lambda$ can be positive or negative and $\mu_{\rm od} $ is bounded, 
taking only values $-0.0197 \,\roughly< \, \mu_{\rm od}\,\roughly< \, 0.0273$. Note that $m_{\rm od}^2$ vanishes when $\xi' \approx -2.31$, and also when $|\xi'| \to \infty$. We comment on the possible phenomenological consequences in Section 6.


\newsec{Away from $U(1)$ flatness}

In our analysis so far we have assumed that 
$h \ll g_{1,4} \ll 1$. This hierarchy forces the minimum of the scalar potential to be close to the D-flat directions. One 
might ask what happens if the condition $h \ll g_1$ is relaxed so that the minimum of the potential moves away from the $U(1)$ flat directions.  Another assumption we made was that the mass $m$ of the messenger fields was large enough that the potential was minimized when the vevs of the messengers vanished. We have not yet made it clear what ``large enough'' means. 
In this section, we explore these two issues.

\subsec{How big is big? (without being too big)}

Let us for simplicity study the 4-1 model and a single pair of messenger fields with $U(1)$ charges $\pm 1$. We do not include an FI term here.
As in Section 2, we rescale all fields $\phi \to \Lambda\, h^{-1/5} \tilde \phi$. The potential can then be written as
\eqn\Vtot{
V = h^{6/5}\, \Lambda^4
\left( V_F 
+ \widetilde{m}^2 \big( |\tilde{L}_+|^2 + |\tilde{L}_-|^2 \big)
+ \frac{1}{\epsilon_1} V_D^{U(1)}
+ \frac{1}{\epsilon_4} V_D^{SU(4)} 
\right) \,  , 
}
where $V_F$ is the F-term potential of the pure 4-1 model and we have introduced dimensionless parameters
\eqn\tildes{
  \widetilde{m}^2 = \frac{m^2}{\Lambda^2\, h^{8/5}} \, ,\quad~~
  \epsilon_{1,4} = \frac{h^2}{g_{1,4}^2} \, . 
}
The $U(1)$ D-term potential
includes the messengers and is
\eqn\dtmsg{
   V_D^{U(1)} = \frac{1}{8}
  \big(d + |\tilde{L}_+|^2 - |\tilde{L}_-|^2\big)^2   \, .
}
where $d= 4|\tilde{S}|^2 - 3 |\tilde{F}|^2 - | \widetilde{\bar F}|^2 + 2 |\tilde{A}|^2$ is the 
$U(1)$ D-term of the pure 4-1 model.

We will consider the limit $\epsilon_4 \to 0$ in which the minimum of the potential is located on the $SU(4)$ D-flat directions. Our job is then to minimize the potential
\eqn\VtotB{
V = h^{6/5}\, \Lambda^4
\left[ V_F 
+ \widetilde{m}^2 \big( |\tilde{L}_+|^2 + |\tilde{L}_-|^2 \big)
+  \frac{1}{8 \epsilon_1}
  \big(d + |\tilde{L}_+|^2 - |\tilde{L}_-|^2\big)^2   
\right] \, 
}
on the $SU(4)$ D-flat directions, where $d=2a^2 - 4b^2 +4c^2$ and $V_F$ is given in \VF .

If we take the limit $\epsilon_1 \to 0$, we must impose the $U(1)$ D-flatness condition $V_D^{U(1)} =0$, and the messenger mass term forces the minimum of $V$ to be at $\langle \tilde L_\pm \rangle = 0$. This in turn enforces the D-flatness condition of the pure 4-1 model, namely $d=0$. The minimization of the remaining potential $V_F$ was the calculation of Section 2.

Let us now consider finite $\epsilon_1 > 0$. From the point of view of the messenger fields, the minimization problem is simply 
SQED with gauge coupling $\tilde{g}_1^2=1/\epsilon_1$ and a Fayet-Iliopoulos term $d$. The role of the FI term is played by the ``distance'' $d$ away from the $U(1)$ flat directions. Extremizing \VtotB, we see that the minimum is located at $\langle \tilde L_\pm \rangle = 0$ when $|d| < 4 \tilde{m}^2 \epsilon_1$, 
and away from the origin otherwise. 
To avoid $\langle \tilde L_\pm \rangle \neq 0$, we must assume that  $m$ satisfies 
$m^2 \roughly>\, g_1^2 \, h^{-2/5} \Lambda^2\,\, |d|/4$.
This is what we mean by $m^2$ being ``large enough''.

However, we do not want to have $m^2$ too big, since we want to be able to integrate out the Higgsed vector fields while keeping the messengers in the resulting effective low-energy theory. The masses of the Higgsed vectors are of order $m_{V_{1,4}}^2 \sim g_{1,4}^2 v^2$ with $v \sim h^{-1/5} \Lambda$. Noting that the lower bound on $m^2$ found in the previous paragraph can be written $m_{V_1}^2 d/4$, we can express the resulting conditions on $m^2$ as the inequality 
\eqn\mcond{
  m_{V_1}^2\, d/4~\,\roughly< ~\, m^2 \,\ll\, m_{V_{1,4}}^2\, .
}
Clearly this can only be satisfied if $d/4 \ll 1$. 

The SUSY-split messenger masses $m^2 \pm m_d^2$ come from the cross-terms in the D-term potential: from \VtotB\ we find $m_d^2 = d/(4\epsilon_1) \, \Lambda^2 h^{8/5} \sim (d/4)\, m^2_{V_1}$. 
 Thus the condition that sends the messengers to the origin, 
$m_{V_1}^2\, d/4~\,\roughly< ~\, m^2$, also ensures that the messenger masses do not become tachyonic.

Let us end this subsection with an example.
Setting $\langle \tilde L_\pm \rangle = 0$, it is 
easy to minimize the potential \VtotB\ for general values of $\epsilon_1$ and compute the corresponding value of $d$ at the minimum. If, in particular, we assume that $\epsilon_1$ is small and expand to linear order, then the minimum is located at
\eqn\epsmin{
(a,b,c)=(1.492 + 0.106\, \epsilon_1,~ 1.102 - 0.0708\,  \epsilon_1,~ 
0.3182 + 0.0554\,  \epsilon_1)\, .
}
These values give $d = 1.396\, \epsilon_1$ and $V_{\rm min} = (2.22-0.244 \epsilon_1) h^{6/5}\, \Lambda^4$. 
The SUSY split masses are
$m_d^2 = d/(4 \epsilon_1) \, \Lambda^2 h^{8/5}= 0.349 \, \Lambda^2 h^{8/5}$. These results agree with what we found in Section 3, see eqs.~\Vnew\ and \mUone . 
The result that $d \sim O(\epsilon_1)$ justifies setting $\langle  \tilde L_\pm \rangle = 0$ provided that $m^2$ satisfies \mcond.

\subsec{Bound on the FI-term}
Let us now consider the same setup as in the previous subsection, but with non-vanishing $\xi$. We will again consider the limit $\epsilon_4 \to 0$ so that we minimize the potential 
\VtotB\ on the $SU(4)$ D-flat directions. Now, however, the $U(1)$ D-term potential involves $d=2a^2 - 4b^2 + 4c^2 + \xi'$ with $\xi'$ the rescaled FI-term, $\xi' = \xi \Lambda^{-2} h^{2/5}$. When $\xi'$ is small, it only acts as a 
perturbation on our earlier analysis. The more interesting case is what happens when $\xi' \gg 1$, since this is when the SUSY-breaking mass splittings $m_d^2$ become large. 

We carry out the calculation as in the previous section, by setting $\langle \tilde L_\pm \rangle = 0$ and expanding in $\epsilon_1 \ll 1$. When $\xi'$ becomes large we find to leading order that the minimum of the potential is located at 
$a \sim \xi'^{-1/3}$, 
$b \sim \xi'^{1/2}/2$, and
$c \sim \xi'^{-7/6}$,
 so that $d \sim 0$. The leading order correction in $\epsilon_1 \ll 1$ gives $d = \epsilon_1 \xi'/2$ at the $O(\epsilon_1)$-corrected minimum of the potential. This in turn implies $m_d^2 = \xi'/8$.

As before, setting $\langle  \tilde L_\pm \rangle = 0$ is justified provided that $|d| < 4 \tilde{m}^2 \epsilon_1$. Again we must make sure that this does not force $m^2$ to be larger than the masses of the Higgsed vectors. An analysis shows that for large $\xi'$, there are 8 heavy vectors with masses $O(\xi')$ and 5 lighter vectors with masses $O(\xi'^{-2/3})$. The latter obviously place the stricter bounds on $m^2$. Restoring the scales, the conditions $m_d^2 < m^2 \ll m_V^2$ can be written
$\Lambda^2 h^{8/5} \xi'/8< m^2 \ll g^2 \Lambda^2 h^{-2/5} \xi'^{-2/3}$ or
\eqn\xim{
\frac{\xi'}{8} 
<  \frac{1}{h^{8/5}}\frac{m^2}{\Lambda^2}  
\ll  \frac{1}{\epsilon}\, \frac{1}{\xi'^{2/3}} \qquad {\rm for\,\, large}~\xi^\prime .
}
Here $\epsilon=h^2/g^2$,
where $g^2$ denotes
 quadratic combinations of $g_1$ and $g_4$. A necessary condition is $\epsilon \ll \xi'^{-5/3}$.  Thus if $\xi'$ is very large, the minimum of the potential is forced to be very close to the D-flat directions.
In addition to this we require that $m^2/\Lambda^2$ and $h$ fulfill the bound \xim.


\subsec{Alternative model}

The Semi-Direct Gauge Mediation models studied here and in \SeibergQJ\ have an explicit dimensionful parameter $m$ which is not dynamically generated. This feature might be considered unattractive. The analyses of the previous two subsections clearly show that the superpotential mass term 
$m\, L_+ L_-$ for the messenger fields is needed in order to stabilize the vacuum at the origin $\langle L_\pm \rangle = 0$. What we explore in this subsection is whether an interaction between the 4-1 model and messenger sector can replace the mass term and thus satisfy the purist's dream of a model with only dynamically generated masses.

We noted in Section 6.1 that from the point of view of the messenger sector, the minimization of the potential is exactly that of the Fayet-Iliopolous model with the role of the FI-term played by $d$, the 4-1 model $U(1)$ D-term. 
When $|d| > 4 \tilde{m}^2 \epsilon_1$, the origin $\langle L_\pm \rangle = 0$ remains a local extremum: it is stabilized in one direction, but becomes tachyonic in the other. In particular, if $d > 4 \tilde{m}^2 \epsilon_1>0$, then $L_-$ is tachyonic at the origin while $L_+$ remains stabilized.  What we seek is an interaction that stabilizes $L_-$ in the limit  $m \to 0$.

Let us for simplicity first consider a single pair of messenger fields $L_\pm$. Then there is a simple construction that does the job. Set $q = 2$ and take the R-charges of $L_\pm$ to be $4$ and $-2$. Then $S (L_-)^2$ is gauge-invariant with R-charge 2. The gauge $U(1)$ and $U(1)_R$ remain anomaly free. The superpotential 
\eqn\newW{
  W  = h S \bar F F + 2 \frac{\Lambda^5}{\sqrt{Y}} + t\, S (L_-)^2
}
preserves the R-symmetry. 

Consider now the scalar potential with an F-term potential from \newW\ and the $U(1)$ D-term potential. When $\epsilon_1$ is non-vanishing, the $L_+$ fields are stabilized at the origin. Moreover, when $|t|$ is sufficiently large,  $L_-$ is also stabilized at the origin. With $\langle L_\pm \rangle = 0$ we can then minimize the 4-1 fields. Expanding around the extremum we find in numerical examples that there are no negative eigenvalues of the boson mass matrix, and this verifies that it is indeed a local minimum.

Introducing the extra term in the superpotential allows a runaway direction where SUSY is restored. This is typical for Minimal Gauge Mediation (MGM) models. There are many possible such runaway directions, even for fixed values of $h$ and $t$, so
it is not clear if the lifetime of the metastable vacuum at $\langle L_\pm \rangle = 0$ can be made sufficiently long-lived.

To convert our simple toy model to a more realistic MGM model requires introducing more messenger fields to get a large enough flavor symmetry group. One way to obtain an $SU(5)$ global symmetry group for the messengers is to
arrange them into adjoints of the $SU(5)$; call them $L_+$ and $L_-$ as before. Then the interaction $S\, \tr\,L_-^2$ in the superpotential preserves the global $SU(5)$. 
The minimization problem involves 24 pairs of messengers and can be carried out as before. We find numerically that the potential has a meta-stable vacuum at $\langle L_\pm \rangle=0$.
 
If we re-introduce the messenger mass term $m$ in the superpotential, we have an MGM model which interpolates between the following extreme limits: (a) At $m=0$, the model has a metastable vacuum; (b) As we take $m \to \infty$ the messengers decouple and SUSY is restored in the SSM; (c) At $t=0$, we recover our 4-1 model of Semi-Direct Gauge Mediation.


\newsec{Phenomenology}

Here we comment briefly on the phenomenology of the Semi-Direct Gauge Mediation model with the 4-1 hidden sector.
We leave a more thorough investigation for future work.

Having a $U(1)$ gauge group in the hidden sector has a few advantages. 
For one, the hierarchy $g_1 \ll g_4$ is automatic, since a gauged $U(1)$ is always IR free. This stands in contrast to the model of \SeibergQJ, where it was necessary to have $N_f$ sufficiently large to achieve the appropriate hierarchy of scales. Since we wish to embed the SSM inside the $SU(N_f)$ flavor symmetry, we must take $N_f \geq 5$.  Another advantage of the $U(1)$ gauge group is that it does not lead to problems with Landau poles in the Standard Model; this is a problem that has plagued Direct Mediation models in the past.

Just as in the 3-2 model, the 4-1 model is automatically CP invariant and the R-symmetry is broken.  
To leading order in $F/m_W$, gaugino masses vanish, just as in \SeibergQJ. This is in accord with the results in \KomargodskiJF. 
There can be several contributions to sfermion masses, starting at the two-loop order \PoppitzXW . As in \SeibergQJ, our model has a negative supertrace over the messenger sector. This will give a positive contribution to the sfermion masses  \PoppitzFW. Additionally, the sfermion masses will get contributions from the tree-level diagonal SUSY-split masses and the off-diagonal masses. A more thorough analysis is needed to determine the overall sign of the sfermion masses.

Adding an FI term makes all the calculated quantities depend on $\xi'=\xi \,h^{2/5}/\Lambda^2$. Gaugino masses still vanish to leading order. The supertrace over the messenger sector remains negative, and decreases monotonically with increasing $\xi'$. Interestingly, the off-diagonal masses coming from the one-loop corrections can now be either positive or negative, and more importantly they are bounded both from above and below. Within the regime of validity of our calculations, it seems that one would be able to tune the supertrace to be large (and negative) while making the off-diagonal masses small. This may help make the sfermion masses positive, but a more detailed analysis is needed in order to see this.

In order for our model to be phenomenologically viable, we would eventually need to couple the model to gravity. One practical reason is that we need gravitational effects to lift massless states, such as the R-axion \BaggerHH\ and the Goldstino. 
Note, however, that it has recently been argued \KomargodskiPC\ that it is not possible to consistently couple a SUSY theory with a (much smaller than Planck scale) FI term to supergravity.  
However, one of the observations we made in Section 5 was that the vanishing of the (vev of the) D-term of other fields can sometimes play the role of an ``effective FI-term''. It would be interesting to exploit this in model building.


\bigskip
\centerline{\bf Acknowledgments}

We would like to thank Nima Arkani-Hamed, Dan Freedman, David Shih, Yuji Tachikawa, Tomer Volansky  and especially Zohar Komargodski and Nati Seiberg for useful discussions and comments on the manuscript.  
BW is supported by DOE grant DE-FG02-90ER40542 and the Frank and Peggy Taplin Membership at the Institute for Advanced Study.
HE is supported by NSF grant PHY-0503584.


\appendix{A}{D-flat directions}

We provide here the basic ingredients needed to construct the D-flatness conditions and solve them.

\subsec{Group theory}

{}Solving for the D-flat directions in this model is made more interesting by the inclusion of the antisymmetric tensor $A_{ij}$. The generators of the two-index antisymmetric representation of $SU(N)$ are
\eqn\asgens{
 \left (T^a_{\sasym}\right)^{~~ij}_{kl}
 = 2  \left (T^a_{\sfund} \right)^{~~[i}_{[k} \delta^{j]}_{l]},
}
where $T^a_{\sfund}$ is a generator of the fundamental of $SU(N)$.
The overall normalization is fixed by requiring that the generators satisfy the algebra. 
With the normalization 
Tr $T^a_{\sfund}T^b_{\sfund} = \delta^{ab}$, the generators of the antisymmetric representation satisfy Tr $T^a_{\sasym}T^b_{\sasym} = (N-2) \delta^{ab}$.

\subsec{D-terms}

The D-term potential $V_D$ of the 4-1 model is given in eq.~\VtotA . We present here the explicit expression for the D-terms. They are
\eqn\DUone{
  D_{U(1)} = 
  \big( 
  q_S\, S^{\dagger} S
  + q_F\, F^{\dagger} F
  + q_{\bar{F}}\, {\bar{F}}^{\dagger}  \bar{F}
  - \frac{1}{2} q_A\, (A^{\dagger})^{ij} A_{ij}
  \big) \, ,
}
\eqn\DSUfour{
\eqalign{
  D_{SU(4)}^a 
  &= 
  \Big( 
  (F^{\dagger})^i (T^a_{\sfund})_i^{~j} F_j
  -  \bar{F}^i  (T^a_{\sfund})_i^{~j} ({\bar{F}}^{\dagger})_j
  - \frac{1}{2} (A^{\dagger})^{ij} (T^a_{\sasym})_{i j}^{~~kl} A_{kl}
  \Big) \, .
  }
  }
The factor $1/2$ for the antisymmetric field is the correct normalization of the K\"ahler term, 
$-\frac{1}{2}(A^\dagger)^{ij} (e^V)_{ij}^{~~kl} A_{kl}$.

Let us write out the $SU(4)$-term for the anti-symmetric field $A_{ij}$ explicitly. We have
\eqn\ad{
(A^{\dagger})^{ij} \left (T^a_{\sasym}\right)^{~~kl}_{ij} A_{kl} 
~=~ 
2 (A^{\dagger})^{ij}  
\left (T^a_{\sfund}\right)^{~~[k}_{[i} \delta^{l]}_{j]}
A_{kl} 
~=~ 
2 (A^{\dagger})^{ik} (T_{\sfund}^a)^{~j}_i  A_{jk}.
}
Thus, the total $SU(4)$ D-term is
\eqn\dsu{
D_{SU(4)}^a 
  = (T^a_{\sfund})_i^{~j} \left [ F^{\dagger i} F_j  - \bar F^i \bar F^\dagger_j - (A^{\dagger})^{ik} A_{jk}\right ] ,
}
which implies that the $SU(4)$ D-flatness condition is
\eqn\dsuii{
 F^{\dagger i} F_j  - \bar F^i \bar F^\dagger_j + (A^{\dagger}A)^{~i}_j= c_0\, \delta^{~i}_{j} 
}
for some complex number $c_0$.

\subsec{Solving the $SU(4)$ D-flatness conditions}

Using $SU(4)$ gauge symmetry, the vevs of the anti-symmetric 4 $\times 4$ matrix $A$ can be brought to the block diagonal form ${\rm diag} (a\, i\, \sigma_2, a'\, i\, \sigma_2)$. Generically, $a \ne a'$, and this then leaves an unbroken $SU(2) \times SU(2)$ subgroup which we can use to rotate the vevs of $F$ to the form $F^T = (f_1,0,f_3,0)$. Let $\bar F = (e_1,e_2,e_3,e_4)$. The D-flatness condition \dsuii\ imposes the constraints $e_2=e_4=0$, $|f_1|=|e_1|$, $|f_3|=|e_3|$ and $|a'| = |a|$.

The group element $U = {\rm diag}(e^{i\phi},1,1,e^{-i\phi})$ of $SU(4)$ can be used to rotate the phases of $a$ and $a'$ so that $a=a'$. With $A={\rm diag} (a\, i\, \sigma_2, a\, i\, \sigma_2)$ there is a larger subgroup of $SU(4)$ which leaves $A$ invariant, namely $Sp(4)$. An element of this group can now be used to rotate $F$ to the form $F^T = (b,0,0,0)$ with $b$ real and non-negative.
The D-term condition \dsuii\ then implies that $e_3=0$, so that $\bar F = (b \, e^{i\phi_b}, 0,0,0)$.

In addition to the local $U(1)$, the theory has a global $U(1)$ symmetry as well as a global $U(1)_R$. 
Using these three $U(1)$'s, we can  set $\phi_b = 0$ and also make $a$ real and non-negative, without introducing new phases in $F$. Thus, without loss of generality, we can parameterize the $SU(4)$ D-flat directions by
\eqn\aisAppFin{
A = \frac{a}{\sqrt{2}} \left ( \matrix{ i\,\sigma_2 & 0 \cr 0 & i \, \sigma_2} \right )\,,\quad
F = \bar F^T=  \left ( \matrix{ b  \cr 0 \cr 0 \cr 0} \right ) \, ,\quad 
 S= c\, e^{i\phi_c} \, , 
}
with real positive numbers $a,b,c > 0$. The $1/\sqrt{2}$ is included in $A$ for later convenience. Eq.~\aisAppFin\ is the result quoted in the main text, see \aisFin .

Note that when messenger fields are added with $U(1)$ gauge charges $\pm q$,  
there is an additional (anomaly-free) global $U(1)$ symmetry under which the 4-1 fields are neutral but the messengers have charges $\pm q'$. If we allow the messengers to acquire vevs, $v_\pm e^{i\phi_\pm}$ (with $v_\pm \ge 0$), then the new global $U(1)$ can be used to set $\phi_- = 0$. This does not interfere with the result \aisAppFin.

\appendix{B}{$SU(4)$ generators}

The 15 generators of $SU(4)$, $T^a = \frac{1}{\sqrt{2}} t^a$, are constructed in analogy with the Gell-Mann matrices. We use the following basis:
\eqn\sufourGen{
{\hesmallrm
 \eqalign{
 & t^1 = \left ( \matrix{ 
 0 & 1 & 0 & 0 \cr
 1 & 0 & 0 & 0 \cr
 0 & 0 & 0 & 0 \cr
 0 & 0 & 0 & 0 
 } \right ) ,
~~~~~~
t^2 = \left ( \matrix{ 
 0 & -i & 0 & 0 \cr
 i & 0 & 0 & 0 \cr
 0 & 0 & 0 & 0 \cr
 0 & 0 & 0 & 0
 } \right ) , 
~~~~
t^3 = \left ( \matrix{ 
 1 & 0 & 0 & 0 \cr
 0 & -1 & 0 & 0 \cr
 0 & 0 & 0 & 0 \cr
 0 & 0 & 0 & 0
 } \right ) , 
~~~~
t^4 = \left ( \matrix{ 
 0 & 0 & 1 & 0 \cr
 0 & 0 & 0 & 0 \cr
 1 & 0 & 0 & 0 \cr
 0 & 0 & 0 & 0
 } \right ) , 
\cr
&
t^5 = \left ( \matrix{ 
 0 & 0 & -i & 0 \cr
 0 & 0 & 0 & 0 \cr
 i & 0 & 0 & 0 \cr
 0 & 0 & 0 & 0
 } \right ) ,  
 ~~~
t^6 = \left ( \matrix{ 
 0 & 0 & 0 & 0 \cr
 0 & 0 & 1 & 0 \cr
 0 & 1 & 0 & 0 \cr
 0 & 0 & 0 & 0
 } \right ) ,  
~~~~~~t^7 = \left ( \matrix{ 
 0 & 0 & 0 & 0 \cr
 0 & 0 & -i & 0 \cr
 0 & i & 0 & 0 \cr
 0 & 0 & 0 & 0
 } \right ) ,   
 ~~~~~
t^8 = \frac{1}{\sqrt{3}}\left ( \matrix{ 
  1 & 0 & 0 & 0 \cr
 0 & 1 & 0 & 0 \cr
 0 & 0 & -2 & 0 \cr
 0 & 0 & 0 & 0 } \right ) ,  
~~~\cr
&
t^9 = \left ( \matrix{ 
 0 & 0 & 0 & 1 \cr
 0 & 0 & 0 & 0 \cr
 0 & 0 & 0 & 0 \cr
 1 & 0 & 0 & 0
 } \right ) ,
 ~~~~~~
 t^{10} = \left ( \matrix{ 
 0 & 0 & 0 & -i \cr
 0 & 0 & 0 & 0 \cr
 0 & 0 & 0 & 0 \cr
 i & 0 & 0 & 0
 } \right ) ,  
 ~~~
t^{11} = \left ( \matrix{ 
 0 & 0 & 0 & 0 \cr
 0 & 0 & 0 & 1 \cr
 0 & 0 & 0 & 0 \cr
 0 & 1 & 0 & 0
 } \right ) ,  
  ~~~
t^{12} = \left ( \matrix{ 
 0 & 0 & 0 & 0 \cr
 0 & 0 & 0 & -i \cr
 0 & 0 & 0 & 0 \cr
 0 & i & 0 & 0
 } \right ) ,  \cr
&
t^{13} = \left ( \matrix{ 
 0 & 0 & 0 & 0 \cr
 0 & 0 & 0 & 0 \cr
 0 & 0 & 0 & 1 \cr
 0 & 0 & 1 & 0
 } \right ) ,  
  ~~~
 t^{14} = \left ( \matrix{ 
 0 & 0 & 0 & 0 \cr
 0 & 0 & 0 & 0 \cr
 0 & 0 & 0 & -i \cr
 0 & 0 & i & 0
 } \right ) ,  
  ~~~
 t^{15} = \frac{1}{\sqrt{6}} \left ( \matrix{ 
 1 & 0 & 0 & 0 \cr
 0 & 1 & 0 & 0 \cr
 0 & 0 & 1 & 0 \cr
 0 & 0 & 0 & -3
 } \right ) .
 }
 }}
 The generators are normalized such that $\tr\, T^a T^b = \delta^{ab}$. The Cartan generators are $T^3$, $T^8$ and $T^{15}$.

On the $SU(4)$ D-flat directions, $F$, $\bar{F}$ and $A$ have vevs \aisAppFin\ which (when $a,b$ are both non-vanishing)  break all generators except $T^{13}$, $T^{14}$, and $\frac{1}{\sqrt{3}}(T^8 -\sqrt{2}\,T^{15})$. Thus the $SU(4)$ is broken to the $SU(2)$ subgroup generated by the 3 unbroken generators.

In our analysis in Section 3.2 we have to remove the 3 unbroken generators when integrating out the Higgsed vector multiplets. The 12 remaining broken generators of $SU(4)$ are $\tilde{T}^a$, $a=1,\dots, 12$, where $\tilde{T}^a = T^a$ when $a \ne 8$ and $\tilde{T}^8 = \frac{1}{\sqrt{3}}(\sqrt{2}\,  T^8 + T^{15} )$. The $U(1)$ gauge group is also broken.

\listrefs
\end